\documentstyle[twoside,fleqn,espcrc2,oldlfont,epsf]{article}
\newcommand{\AmS}{{\protect\the\textfont2
  A\kern-.1667em\lower.5ex\hbox{M}\kern-.125emS}}
\newcommand{\n}{\hspace*{-2.5mm}}
\newcommand{\as}{\alpha_{\rm s}}
\newcommand{\ylab}{y_{\,\rm lab}}
\newcommand{\ycms}{y_{\rm CM}}
\newcommand{\lsim}{\mathrel{\mathop
  {\hbox{\lower0.5ex\hbox{$\sim$}\kern-0.8em\lower-0.7ex\hbox{$<$}}}}}
\newcommand{\gsim}{\mathrel{\mathop
  {\hbox{\lower0.5ex\hbox{$\sim$}\kern-0.8em\lower-0.7ex\hbox{$>$}}}}}
\newcommand{\ms}{$\overline{\rm MS}$}
\renewcommand{\pt}{p_{\rm T}}
\renewcommand{\O}{{\cal O}}

\title{\rightline {hep-ph/9407360}
Inclusive photoproduction of hadrons at HERA\thanks{To appear in the
{\it Proceedings of the Workshop on Two-Photon Physics at LEP and HERA---Status
of Data and Models}, Lund, Sweden, May 26--28, 1994.}}
\author{Bernd A. Kniehl\\
II. Institut f\"ur Theoretische Physik,$\!$\thanks{Supported
by Bundesministerium f\"ur Forschung und Technologie, Bonn, Germany,
under Contract 05~6~HH 93P~(5)
and by EEC Program {\it Human Capital and Mobility} through Network
{\it Physics at High Energy Colliders} under Contract
CHRX-CT93-0357 (DG12~COMA).}
Universit\"at Hamburg,\\
Luruper Chaussee 149, 22761 Hamburg, Germany}

\begin{document}

\begin{abstract}

We study inclusive photoproduction of hadrons at HERA via both direct and
resolved photons to next-to-leading order (NLO) in the QCD-improved parton
model.
We predict transverse-momentum ($\pt$) and rapidity ($\ylab$) distributions
and compare them with recent data by the H1 Collaboration.
To estimate the theoretical error of the predictions, we investigate their
dependence on the choice of renormalization and factorization scales,
subtraction scheme, and parton density functions (PDF).
We also introduce a new NLO set of fragmentation functions for charged pions
and kaons, which is extracted from $e^+e^-$ data at $\sqrt s=29$~GeV, and
confront it with the H1 data.
This enables us to test quantitatively the QCD-improved parton model and,
in particular, the factorization theorem at the quantum level.

\end{abstract}

\maketitle

\section{Introduction}

The advent of high-statistics data on inclusive photoproduction of charged
hadrons, $h$, in the $ep$ scattering experiments performed by the H1
\cite{ahm,abt} and ZEUS \cite{sta} Collaborations at HERA offers the
challenging opportunity to test the QCD-improved parton model quantitatively
at centre-of-mass (c.m.) energies ($\sqrt s$) that exceed those achieved at
previous fixed-target experiments \cite{aps} by more than an order of magnitude.
According to present HERA conditions, $E_e=26.7$~GeV electrons collide with
$E_p=820$~GeV protons in the laboratory frame, so that $\sqrt s=296$~GeV is
available in the c.m.\ frame.
Among the experimentalists it has become customary to take the rapidity,
$\ylab$, of hadrons travelling in the proton direction to be positive.
The c.m.\ rapidity, $\ycms$, is related to $\ylab$ by
\begin{equation}
\ycms=\ylab-{1\over2}\ln{E_p\over E_e}.
\end{equation}

In photoproduction, the electron beam acts like a source of quasi-real
photons, so that HERA is operated effectively as a $\gamma p$ collider.
The appropriate events may be discriminated from deep-inelastic-scattering
events by electron tagging or anti-tagging \cite{kes}.
The photon flux is well approximated by the Weizs\"acker-Williams formula
\cite{kes},
\begin{eqnarray}
\label{wwa}
f_{\gamma/e}(x)&\n=\n&{\alpha\over2\pi}\left[{1+(1-x)^2\over x}
\ln{Q_{\rm max}^2\over Q_{\rm min}^2}\right.\nonumber\\
&\n+\n&\left.
2m_e^2x\left({1\over Q_{\rm max}^2}-{1\over Q_{\rm min}^2}\right)\right],
\end{eqnarray}
where $x=E_\gamma/E_e$, $Q_{\rm min}^2=m_e^2x^2/(1-x)$, and
$Q_{\rm max}^2=0.01$~GeV$^2$ for tagged events at H1 \cite{ros}.
The importance of the second term in Eq.~(\ref{wwa}), which has been
known for a couple of decades \cite{kes}, for such small values
of $Q_{\rm max}^2$ has been re-emphasized recently in Ref.~\cite{fri}.
The $ep\to h+X$ and $\gamma p\to h+X$ cross sections are related by
\begin{eqnarray}
E_h{d^3\sigma(ep\to h+X)\over d^3p_h}
&\n=\n&\int_{x_{\rm min}}^1dx\,f_{\gamma/e}(x)\nonumber\\
&\n\times\n&
E_h{d^3\sigma(\gamma p\to h+X)\over d^3p_h},
\end{eqnarray}
where $x_{\rm min}=\pt\exp(-\ycms)/[\sqrt s-\pt\exp(\ycms)]$.
In the following, we shall impose $0.3<x<0.7$ to be in conformity with the H1
event-selection criteria.

It is well known that $\gamma p\to h+X$ proceeds via two distinct mechanisms.
The photon can interact either directly with the partons originating from the
proton (direct photoproduction) or via its quark and gluon content (resolved
photoproduction).
Both contributions are formally of the same order in the perturbative expansion.
Leaving aside the proton PDF, $G_{b/p}(x,M_p^2)$, and the fragmentation
functions, $D_{h/c}(x,M_h^2)$, which represent common factors,
the LO cross sections are of $\O(\alpha\as)$ in both cases.
In the case of the resolved mechanism, this may be understood by observing 
that the $ab\to cd$ cross sections, which are of $\O(\as^2)$, get dressed
by photon PDF, $G_{a/\gamma}(x,M_\gamma^2)$, whose leading terms are of the
form $\alpha\ln(M_\gamma^2/\Lambda^2)\propto\alpha\as$.
Here, $a,b,c,d$ denote quarks and gluons.
In fact, the two mechanisms compete with each other also numerically.
Resolved photoproduction dominates at small $\pt$ and positive $\ylab$,
while direct photoproduction wins out at large $\pt$ and negative $\ylab$.

The LO calculation suffers from significant theoretical uncertainties
connected with the freedom in the choice of the renormalization scale, $\mu$,
of $\as(\mu^2)$ and the factorization scales, $M_\gamma,M_p,M_h$.\footnote{%
The LO cross section of direct photoproduction does not depend on $M_\gamma$.}
In order to obtain reliable predictions, it is indispensable to proceed to NLO.
In this contribution, we shall report the results of analyses where this
problem has been tackled \cite{bor,kni,gor,gre}.

\section{Formalism}

Prior to presenting the numerical results, we shall introduce the NLO 
formalism.
We shall first consider resolved photoproduction, which is more involved.
Starting out from the well-known LO cross section of $\gamma p\to h+X$
\cite{bor},
one needs to include the NLO corrections, $K_{ab\to c}$, to the
hard-scattering cross sections, to substitute the two-loop formula for $\as$,
and to endow $G_{a/\gamma}$, $G_{b/p}$, and $D_{h/c}$ with NLO evolution.
This leads to
\begin{eqnarray}
\label{res}
&&\hspace*{-7.mm}E_h{d^3\sigma(\gamma p\to h+X)\over d^3p_h}\nonumber\\
&&\hspace*{-7.mm}{}=\sum_{a,b,c}\int dx_\gamma dx_p{dx_h\over x_h^2}\,
G_{a/\gamma}(x_\gamma,M_\gamma^2)G_{b/p}(x_p,M_p^2)\nonumber\\
&&\hspace*{-7.mm}{}\times D_{h/c}(x_h,M_h^2){1\over\pi s}
\left[{1\over v}\,{d\sigma_{ab\to c}^0\over dv}(s,v;\mu^2)
\delta(1-w)\right.\cr
&&\hspace*{-7.mm}{}+\left.\vphantom{d\sigma_{ab\to c}^0\over dv}
{\as(\mu^2)\over2\pi}
K_{ab\to c}(s,v,w;\mu^2,M_\gamma^2,M_p^2,M_h^2)\right],
\end{eqnarray}
where $v=1+t/s$ and $w=-u/(s+t)$, with $s=(p_a+p_b)^2$, $t=(p_a-p_c)^2$,
and $u=(p_b-p_c)^2$ being the Mandelstam variables at the parton level.
The indices $a,b,c$ run over the gluon and $N_F$ active flavours of quarks
and antiquarks.
In the process at hand, we take $N_F=4$.
The $K_{ab\to c}$ functions may be found in Ref.~\cite{ave} for $M_\gamma=M_p$.
This restriction was relaxed in Ref.~\cite{kni}.

The NLO cross section of direct photoproduction emerges from Eq.~(\ref{res})
by substituting $G_{a/\gamma}(x_\gamma,M_\gamma^2)=\delta(1-x_\gamma)$,
replacing $d\sigma_{ab\to c}/dv$ and $K_{ab\to c}$ by
$d\sigma_{\gamma b\to c}/dv$ and $K_{\gamma b\to c}$, respectively,
and omitting the sum over $a$.
The $K_{\gamma b\to c}$ functions were first derived in Ref.~\cite{aur}
setting $M_\gamma=M_p=M_h$ and taking the spin-average for incoming
photons and gluons to be 1/2.
In Ref.~\cite{kni}, the scales were disentangled and the spin-average
convention was converted to the \ms\ scheme, i.e., to be $1/(n-2)$, with
$n$ being the dimensionality of space-time.
Analytic expressions for the $K_{\gamma b\to c}$ functions are listed in
Ref.~\cite{gor}.

In the evaluation of the $K$ functions, one encounters ultraviolet,
infrared, and collinear singularities, which, in dimensional regularization,
are extracted as poles in $\epsilon=2-n/2$.
The ultraviolet singularities appear as $1/\epsilon$ poles in the
virtual corrections and are cancelled by the \ms\ counterterms, which
are derived from the LO cross sections in the usual way.
There remains an explicit $\mu$ dependence in the $K$ functions.
The infrared singularities, whose leading terms are proportional to
$1/\epsilon^2$, cancel between the virtual and real corrections.
After that, one is left with collinear singularities connected with
the incoming and outgoing legs.
They are proportional to $1/\epsilon$ with coefficients that
are uniquely determined by the factorization theorem.
These divergences are subtracted at scales $M_\gamma,M_p,M_h$ and absorbed into
the bare photon and proton PDF and fragmentation functions
rendering these functions finite.
In particular, the collinear singularity associated with the incident
photon of the direct process is shifted to the photon PDF of the resolved
process.
In this way the direct and resolved contributions become intermixed at NLO.
Both components depend strongly on $M_\gamma$, while their superposition
is almost insensitive to $M_\gamma$.
On the other hand, the $\mu,M_p,M_h$ dependences are reduced at NLO within
each component separately.

\begin{figure}[t]
\epsfxsize=7.5cm 
\epsfbox[40 324 559 735]{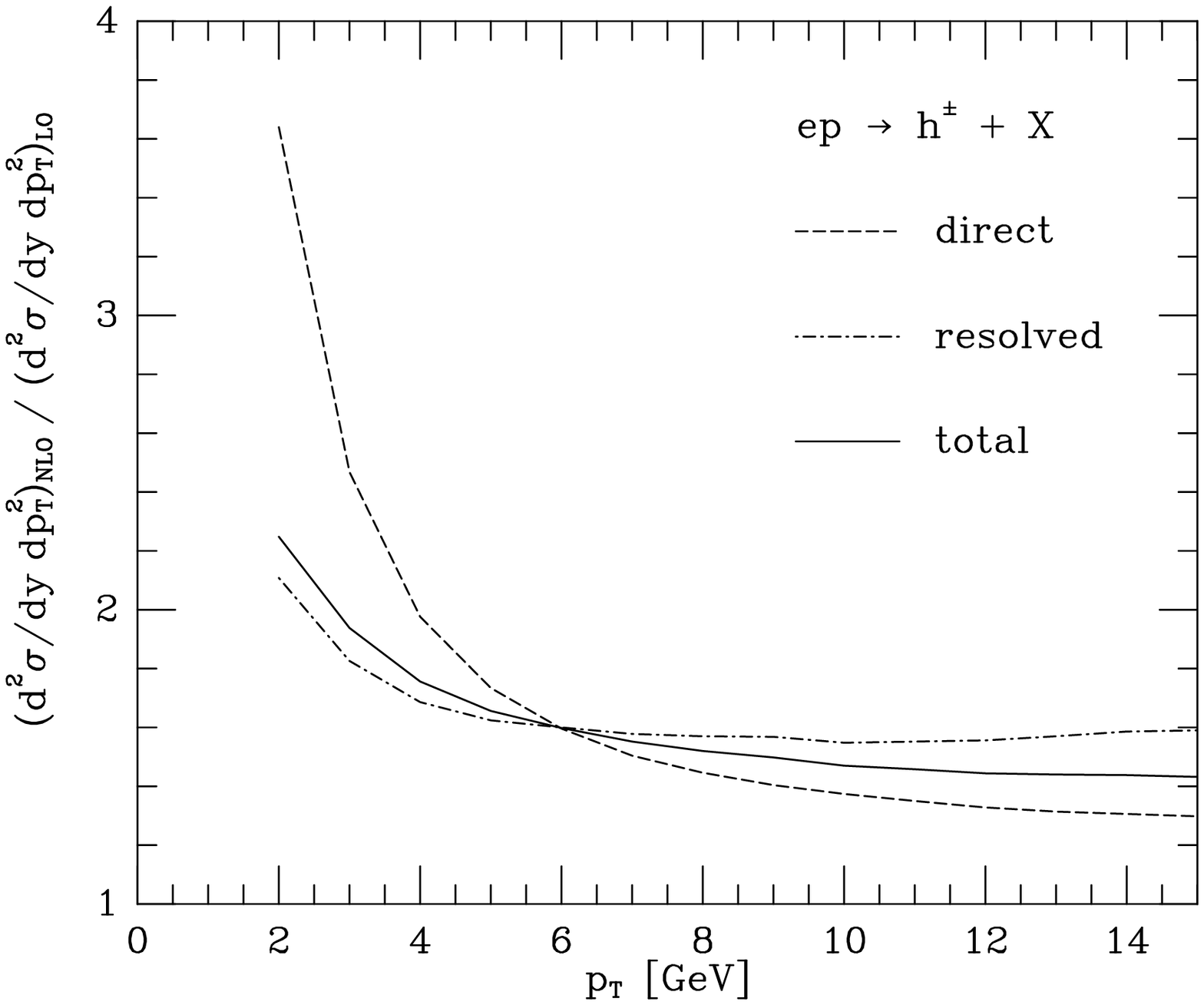}
\vskip-.8cm
\caption{$\pt$ distributions of $ep\to h+X$ via direct (dashed line)
and resolved (dot-dashed line) photons and their superposition (solid line)
averaged over $|\ylab|<1$.
The NLO results are normalized to the respective evaluations with the $K$
functions set to zero}
\label{figone}
\vskip-.8cm
\end{figure}

The subtraction of the collinear divergences is not unique.
In the \ms\ scheme, one subtracts just the $1/\epsilon$ poles along
with the usual combination $\ln(4\pi)-\gamma_E$, where $\gamma_E$ is
Euler's constant.
In principle, one can include also an arbitrary dimensionless function,
$f_{ia}(z)$, where $p_i=zp_a$, in the subtraction related to the transition
of an incoming parton $a$ into parton $i$ under collinear radiation.
Similar functions, $d_{ci}(z)$, with $p_c=zp_i$ can be introduced also for
outgoing partons $c$.
This leads to modifications in the $K$ functions, which are accompanied by
appropriate redefinitions of the $G_{a/\gamma}$, $G_{b/p}$, and $D_{h/c}$
functions.
In the case of the fragmentation functions, one usually sticks to the \ms\
scheme.
In the case of the proton PDF, some authors prefer the so-called
deep-inelastic-scattering (DIS) scheme, which is tailored in such a way that
the NLO structure function $F_2$ for lepton-nucleon DIS assumes a form
similar to the one in the na\"\i ve parton model \cite{die}.
This concept has been adopted also for the photon PDF, where it is called
DIS$_\gamma$ scheme \cite{grv}.
It is apparent that the evaluations of Eq.~(\ref{res}) and the analogous
equation for direct photoproduction in different subtraction schemes differ
by terms beyond NLO.
Such differences also contribute to the theoretical error of the predictions.

\section{Numerical results}

\begin{figure}[t]
\epsfxsize=7.5cm 
\epsfbox[40 324 559 735]{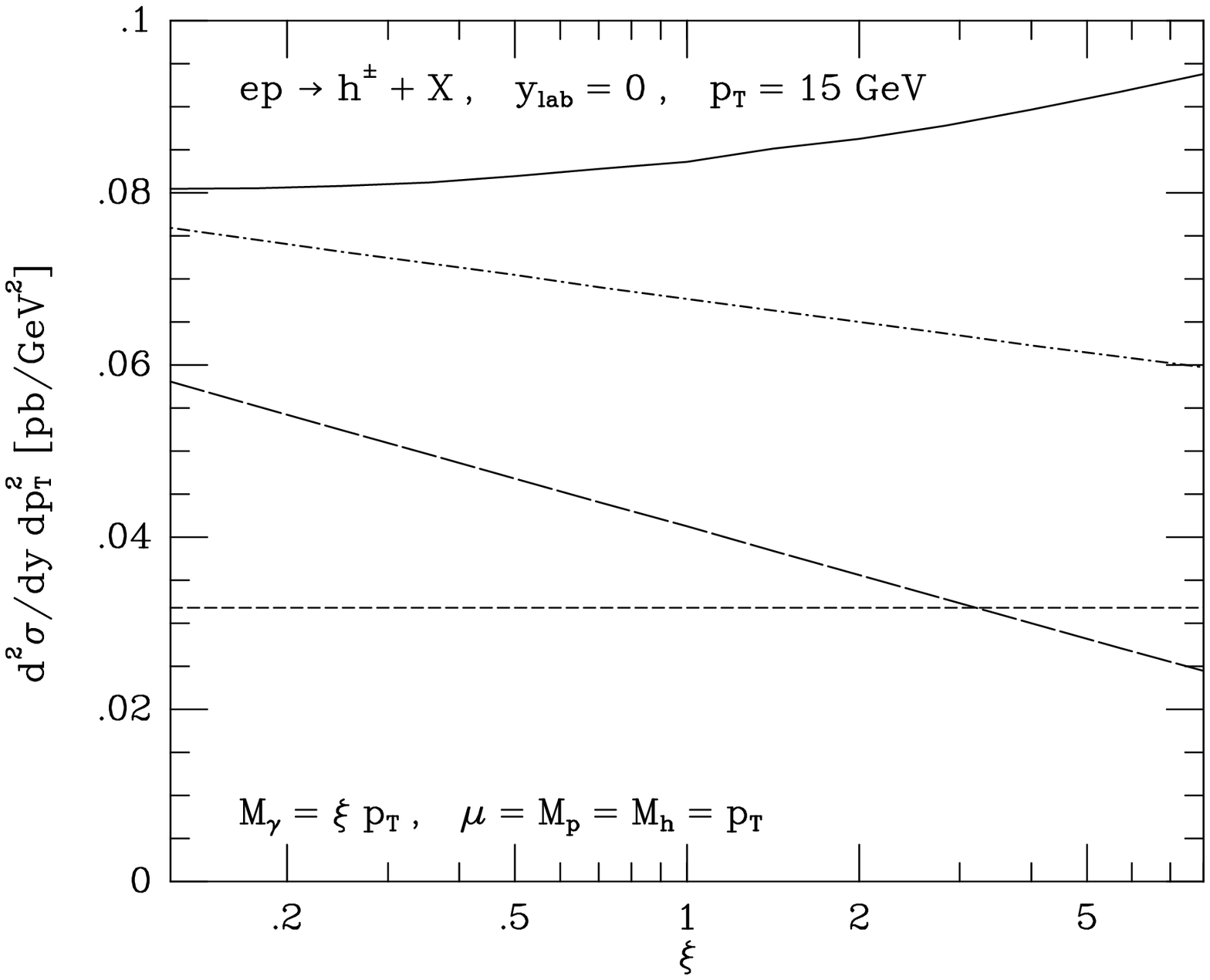}
\vskip-.8cm
\caption{$M_\gamma$ dependence ($M_\gamma=\xi\pt$) of
$d^2\sigma/dy\,d\pt^2$ for $ep\to h+X$ via photoproduction
at $\ylab=0$ and $\pt=15$~GeV.
Starting from the LO direct cross section (dashed line), we add in turn
the NLO corrections to the direct cross section (long-dashed line),
the LO resolved cross section (dot-dashed line),
and the NLO corrections to the resolved cross section (solid line)}
\label{figtwo}
\vskip-.8cm
\end{figure}

We are now in a position to present our numerical results and to assess
their theoretical uncertainties.
Unless stated otherwise, we shall work in the \ms\ scheme with $N_F=4$
active quark flavours, set $\mu=M_\gamma=M_p=M_h=\pt$, and employ
the NLO set of the photon PDF by Gl\"uck, Reya, and Vogt \cite{grv}, which
we translate to the \ms\ scheme,
set D$_0^\prime$ of the proton PDF by Martin, Roberts, and Stirling
\cite{mrs}, and the fragmentation functions by Baier, Engels, and
Petersson \cite{bep}, which were updated and complemented by Anselmino,
Kroll, and Leader \cite{akl}.
These fragmentation functions are LO fits to rather old data of low-energy
$e^+e^-$ scattering and deep-inelastic muon-nucleon scattering.
Although we have tested them in $p\bar p$ scattering \cite{fmb},
they nevertheless imply a possible limitation of our NLO analysis.
In order to eliminate this potential weakness, we have recently generated
a NLO set of fragmentation functions, which are fitted to $e^+e^-$ data
taken $\sqrt s=29$~GeV by the TPC Collaboration at SLAC \cite{bkk}.

\begin{figure}[t]
\epsfxsize=7.5cm 
\epsfbox[40 324 559 735]{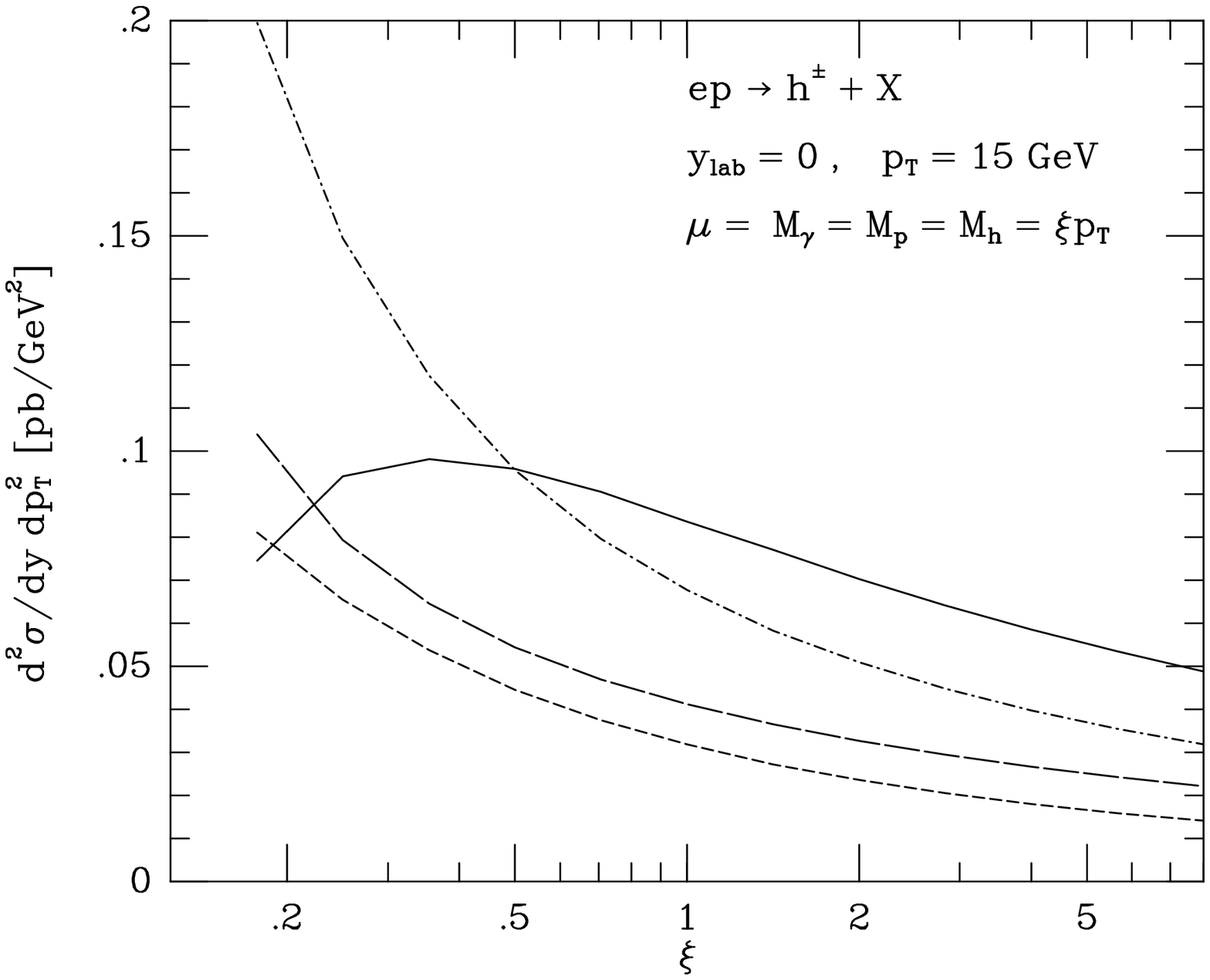}
\vskip-.8cm
\caption{Same as Fig.~\protect\ref{figtwo} for the global scale dependence
($\mu=M_\gamma=M_p=M_h=\xi\pt$)}
\label{figthree}
\vskip-1.cm
\end{figure}

In Fig.~\ref{figone}, we present the ratios of the direct, resolved, and total
$ep$ cross sections at NLO to the respective evaluations with the $K$
functions omitted.
The direct cross section receives significant NLO corrections at very low
$\pt$, where it is much smaller than the resolved one anyway.  
In the region where the QCD-improved parton model is expected to yield
reliable predictions, $\pt\gsim5$~GeV, the increase due to the NLO
corrections is relatively moderate, below 65\%.

As mentioned above,
the relation of the direct and resolved components varies with the
choice of $M_\gamma$ and/or factorization scheme.
If we put $M_\gamma=\xi\pt$ and vary $\xi$, then the
NLO direct part changes logarithmically with $\ln\xi$,
which is compensated partly already by the LO resolved cross section.
This compensation is demonstrated in Fig.~\ref{figtwo}, where we show
$d^2\sigma/dy\,d\pt^2$ for $ep\to h+X$ at $\ylab=0$ and $\pt=15$~GeV
as a function of $\xi$, keeping all other scales fixed at $\pt$.
Of course, the LO direct cross section is independent of $\xi$ (dashed line).
The NLO direct cross section is a decreasing monomial in $\ln\xi$ (long-dashed
line).
As expected, the $\xi$ dependence is significantly reduced when the NLO
direct and LO resolved contributions are combined (dot-dashed line).
A complete cancellation is beyond the scope of finite-order perturbation
theory; there are always uncompensated terms in the next order beyond
current knowledge.
We also show the sum of the NLO direct and NLO resolved cross sections,
which now feebly increases with $\xi$, but stays tolerably constant
over the central range $0.5<\xi<2$, which is usually considered in
connection with scale variations.
In Ref.~\cite{kni}, the factorization-scheme dependence of the NLO calculation
is investigated separately for the photon and proton legs.
In the case of the proton, there is a compensation between the proton
PDF and the $K$ functions within the direct and resolved cross sections,
while in the case of the photon only the sum of the two cross sections
is stable under the change of scheme. 

\begin{figure}[t]
\epsfxsize=7.5cm 
\epsfbox[20 74 556 731]{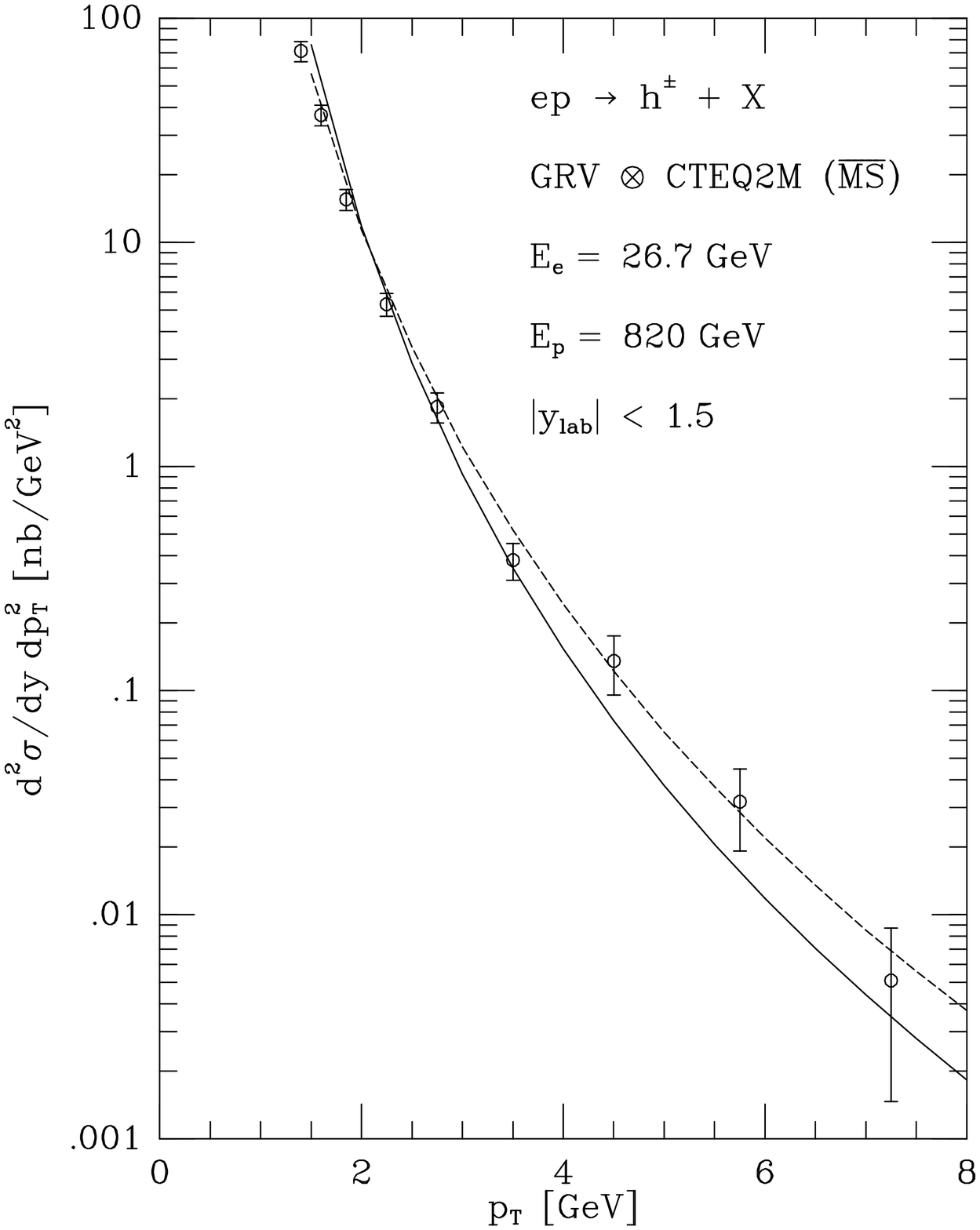}
\vskip-.8cm
\caption{$\pt$ distribution of $ep\to h+X$ via photoproduction at NLO
averaged over $|\ylab|<1.5$.
The dashed (solid) line is evaluated with the LO (NLO) fragmentation
functions of Ref.~\protect\cite{bep,akl} (\protect\cite{bkk})}
\label{figfour}
\vskip-.8cm
\end{figure}

The global scale dependence, i.e., the one generated by setting
$\mu=M_\gamma=M_p=M_h=\xi\pt$ and varying $\xi$, is investigated
in Fig.~\ref{figthree} for $\ylab=0$ and $\pt=15$~GeV.
It is much stronger than the $M_\gamma$ dependence alone.
As expected, the summed NLO cross section (NLO direct plus NLO resolved)
is much more stable than the LO direct one.
>From Fig.~\ref{figthree} we read off a theoretical uncertainty of
20\% for $\pt\ge15$~GeV
due to a global scale variation in the range $0.5<\xi<2$.

In Fig.~\ref{figfour}, we confront our NLO prediction with recent H1 data on
the $\pt$ spectrum averaged over $|\ylab|<1.5$ \cite{abt}.
A similar comparison for the $\ylab$ spectrum may be found in Ref.~\cite{abt}.
Here, we use set CTEQ2M of the proton PDF from Ref.~\cite{bot}. 
The dashed line is evaluated with the old LO fragmentation functions
\cite{bep,akl} and the solid one with our new NLO set \cite{bkk}.
Note that the first evaluation is somewhat inconsistent, since the $M_h$
dependence of the $K$ functions is not compensated, so that the good agreement
with the data is perhaps somewhat fortuitous.

It is interesting to study whether the data already show evidence for the
presence of both the resolved and direct mechanisms.
At $\pt=2$~GeV (8~GeV), 17\% (37\%) of the photoproduction of hadrons proceeds
via the direct mechanism.
At $\pt\approx15$~GeV, the direct mechanism surpasses the resolved one.
Looking at Fig.~\ref{figfour}, the data seem to indicate that both mechanisms
do contribute.
However, in order to make a firm statement, we need to wait for higher
statistics, especially in the upper $\pt$ range.

Another interesting issue is whether the gluon content of the resolved photon
has been established by the data.
At $\pt=2$~GeV (8~GeV), it makes up 49\% (15\%) of the resolved contribution
and 41\% (9\%) of the total one.
Thus, it seems that, in the absence of the gluon density inside the photon,
the theoretical prediction would fall short of the data significantly at low
$\pt$.
Encouraged by this observation, one might go one step further and try to
see if the data favour or disfavour certain sets of photon PDF.
In fact, the available sets differ mainly by the gluon density.
On the other hand, the modern sets of proton PDF \cite{mrs,bot} yield rather
similar results.
It turns out that the data strongly disfavour set III of the photon PDF by
Abramowicz, Charchu\l a, and Levy \cite{lac}, which has a very hard gluon and
leads to a prediction that overshoots the data by 60--150\%.

\smallskip
\noindent
{\bf Acknowledgements.}
I am grateful to G. Kramer for his collaboration in the work presented here
and to the organizers of the
{\it Workshop on Two-Photon Physics at LEP and HERA}
for creating a stimulating atmosphere.

\end{document}